\documentclass[a4paper,12pt]{article}
\usepackage{amssymb,float,amsmath,amsfonts,amsthm,cases,cite}
\usepackage[normalem]{ulem}

\usepackage[pdfencoding=auto]{hyperref}
\usepackage{xcolor}
\usepackage{hyperref}
\definecolor{linkcolor}{HTML}{799B03}
\definecolor{urlcolor}{HTML}{799B03}
\hypersetup{pdfstartview=FitH,linkcolor=linkcolor,urlcolor=urlcolor,colorlinks=true}

\def\be{\begin{equation}\begin{aligned}}
\def\ee{\end{aligned}\end{equation}}

\begin{document}

\vspace{10pt}

\begin{center}
{\LARGE \bf Properties of perturbations in beyond Horndeski theories}

\vspace{20pt}

S. Mironov$^{a,c}$\footnote[1]{sa.mironov\_1@physics.msu.ru},
V. Volkova$^{a,b}$\footnote[2]{volkova.viktoriya@physics.msu.ru}

\vspace{15pt}

$^a$\textit{Institute for Nuclear Research of the Russian Academy of Sciences,\\
60th October Anniversary Prospect, 7a, 117312 Moscow, Russia}\\
\vspace{5pt}

$^b$\textit{Department of Particle Physics and Cosmology, Physics Faculty,\\
M.V. Lomonosov Moscow State University,\\
Vorobjevy Gory, 119991 Moscow, Russia}

$^c$\textit{Institute for Theoretical and Experimental Physics,\\
Bolshaya Cheriomyshkinskaya, 25, 117218 Moscow, Russia}
\end{center}

\vspace{5pt}

\begin{abstract}
We study whether the approach of Deffayet et al. (DPSV) can be adopted for obtaining a derivative part of quadratic action for scalar perturbations in beyond Horndeski theories about homogeneous and isotropic backgrounds. We find that even though the method does remove the second and higher derivatives of metric perturbations from the linearized Galileon equation, in the same manner as in the general Horndeski theory, it gives incorrect result for the quadratic action. We analyse the reasons behind this property and suggest the way of modifying the approach, so that it gives valid results.
\end{abstract}

\section{Introduction and summary}
\label{sec:intro}
One of the notable features of the generalized Galileon theory (Horndeski and beyond\footnote[3]{In this paper, we refer to the scalar field in the theory as Galileon irrespectively of the subclass we discuss.}) is
the ability to violate the Null Energy Condition in a healthy way. The latter fact enables one to
construct the unconventional
cosmological solutions like bouncing Universe
and Genesis.
However, many classical
cosmological models of this kind develop instabilities (ghosts, gradient instabilities, etc.)
at early or late times~\cite{Easson:2011zy, Cai:2012va,  Koehn:2013upa, Battarra:2014tga, Qiu:2015nha, Kobayashi:2015gga, Wan:2015hya, Ijjas:2016tpn}.
Recently, within the ``beyond Horndeski'' theory,
several bouncing and Genesis
cosmological solutions,
which remain stable during the entire classical evolution~\cite{Libanov:2016kfc,Cai:2016thi,Creminelli:2016zwa,Kolevatov:2017voe,Cai:2017dyi}, have been
constructed. In any case, the issue of stability is central for these constructions.

The most straightforward way to study the stability at the linearized level is to choose a parametrization of perturbations, plug it into the action and expand the latter to the second order.
Things simplify in the unitary gauge, where the Galileon perturbation vanishes.
By integrating out the non-dynamical degrees of freedom, one obtains
the quadratic action for the remaining degrees of freedom: two tensor and one scalar (curvature in this case).
This approach has been adopted in a number of works \cite{Deffayet:2010qz, Kobayashi:2010cm,Kobayashi:2011nu,Kobayashi:2016xpl,Kolevatov:2017voe} .

However, there is an alternative method, proposed in Ref.~\cite{Deffayet:2010qz}
for the cubic Galileon theory. We refer to it as the DPSV approach for brevity.
This is a bottom up approach, in which the quadratic action is restored from the linearized equation.
It has been noticed in Ref.~\cite{Deffayet:2010qz} that the Galileon and Einstein equations
in the cubic Galileon theory contain the second derivatives of both metric and
Galileon perturbations. It was suggested to make use of the Einstein equations to eliminate
the second derivatives of the metric from the Galileon field equation, so that the second derivative terms
in the resulting equation involve the Galileon only. In the cubic theory this procedure goes through
at the fully non-linear level. Upon linearizing the resulting Galileon equation one pretends to
reconstruct the derivative part of the quadratic action with a single degree of freedom,
which is the Galileon perturbation in this case. It is worth noting that one drops the first and zeroth derivatives
of both metric and Galileon field in the resulting linearized Galileon equation, having in mind
that high momentum
modes are of interest only.
The advantage of this alternative approach is that it works with the linearized equations, while other conventional methods require a quadratic action to be calculated from the scratch.

Originally, Deffayet et al. have checked in the Appendix of Ref.~\cite{Deffayet:2010qz} that the trick indeed gives correct result in the cubic theory, but the reasons behind the fact that the approach works correctly were not addressed. Since the trick utilized unperturbed Galileon and Einstein equations, there were no suggestions as to whether the trick amounts to a specific gauge choice. However, at the linearized level it does, as shown in a recent work~\cite{Kolevatov:2017dze}.    
In Ref.~\cite{Kolevatov:2017dze} the trick has been studied within the cubic + quartic theory
and was shown to give the derivative part of quadratic action for the Galileon perturbation, which agrees with the corresponding results in the unitary gauge. Quite a detailed analysis of the reasons behind the validity of the trick has been provided.

This paper is sequential to Ref.\cite{Kolevatov:2017dze} and aims to find out whether the DPSV trick works in the beyond Horndeski case.
The motivation for this study is twofold. Firstly, if the trick is
applicable, namely, one can eliminate
the second derivatives of the metric from the linearized Galileon equation, and the
quadratic action for the Galileon perturbation obtained in this way is correct,
it will be legitimate to utilize this method as a cross-check of the unitary gauge result.
Secondly, and most importantly, the DPSV trick significantly simplified the study of wormhole stability~\cite{Rubakov:2016zah, Kolevatov:2016ppi,Evseev:2016ppw}
and might be useful for generalizing the existing results. Our result, however, is as follows.
We find that the DPSV approach does remove higher derivatives of metric perturbations in the
beyond Horndeski case;
however, the result it gives for the quadratic action is incorrect. Thus, the original DPSV trick does not work beyond Horndeski. We discuss the reason behind the trick's break down and suggest the way to restore the correct result.

This paper is organized as follows. We introduce notations and summarise earlier results in Sec.~\ref{sec:Summary} to clarify the issue that is discussed in this paper.
We employ the DPSV trick in the most general Horndeski theory
in Sec.~\ref{sec:DPSV_for_L5} and show that the result for the quadratic action coincides with
that obtained via the unitary gauge method in Ref.~\cite{Kobayashi:2011nu}.
In Sec.\ref{sec:DPSV_for_BH} we attempt to apply the DPSV approach in the beyond Horndeski case
and find that it eliminates higher derivatives of metric perturbations, but the reconstructed
quadratic action does not agree with the result of Ref.\cite{Kolevatov:2017voe}.
We give the explanation of the latter fact in Sec.\ref{sec:alpha_beta_for_BH}.

\section{Review of existing results}
\label{sec:Summary}
The most general Lagrangian for the generalized Galileons
reads:
\begin{equation*}
\mathcal{L} = \mathcal{L}_3 + \mathcal{L}_4 + \mathcal{L}_5 + \mathcal{L_{BH}},
\end{equation*}
where
\begin{subequations}
\label{lagrangian}
    \begin{align}
    \label{L3}
    &\mathcal{L}_3=F(\pi,X) + K(\pi,X)\Box\pi,\\
    \label{L4}
    &\mathcal{L}_4=-G_4(\pi,X)R+2G_{4X}(\pi,X)\left[\left(\Box\pi\right)^2-\pi_{;\mu\nu}\pi^{;\mu\nu}\right],\\
    \label{L5}
    &\mathcal{L}_5=G_5(\pi,X)G^{\mu\nu}\pi_{;\mu\nu}+\frac{1}{3}G_{5X}\left[\left(\Box\pi\right)^3-3\Box\pi\pi_{;\mu\nu}\pi^{;\mu\nu}+2\pi_{;\mu\nu}\pi^{;\mu\rho}\pi_{;\rho}^{\;\;\nu}\right],\\
    \label{BH}
    &\mathcal{L_{BH}}=F_4(\pi,X)\epsilon^{\mu\nu\rho}_{\quad\;\sigma}\epsilon^{\mu'\nu'\rho'\sigma}\pi_{,\mu}\pi_{,\mu'}\pi_{;\nu\nu'}\pi_{;\rho\rho'}
    \\\nonumber&\qquad+F_5(\pi,X)\epsilon^{\mu\nu\rho\sigma}\epsilon^{\mu'\nu'\rho'\sigma'}\pi_{,\mu}\pi_{,\mu'}\pi_{;\nu\nu'}\pi_{;\rho\rho'}\pi_{;\sigma\sigma'}.
    \end{align}
\end{subequations}
Here $\pi$ is the Galileon field, $X=g^{\mu\nu}\nabla_{\mu}\pi
\nabla_{\nu}\pi$, $\pi_{,\mu} = \partial_{\mu}\pi$,
$\pi_{;\mu\nu}=\triangledown_\nu\triangledown_\mu\pi$, $\Box\pi =
g^{\mu\nu}\triangledown_\nu\triangledown_\mu\pi$, $G_{4X}=\partial
G_4/\partial X$, etc. The terms~\eqref{BH} emerge beyond Horndeski,
while others are characteristic of the Horndeski theory. The major
difference between Horndeski and beyond Horndeski theories is the
fact that in the latter, equations of motion are of the third order.
Nevertheless, propagating perturbations still obey second-order
equations of motion and hence are free of Ostrogradski
instabilities~\cite{Gleyzes:2014dya,Langlois:2017mxy}.

In this paper we study the generalized Galileon theory
in a spatially flat cosmological setting.
Instabilities, if any, show up at the level of small perturbations about a background solution.
In what follows we focus on high momenta and frequency modes only, since these are
the most harmful ones. Hence, the most relevant
are the second and higher derivatives of perturbations
in the linearized equations of motion. We use the following notations
in the scalar sector:
\begin{equation}
\label{perturbations}
h_{00}=2\alpha,\quad h_{0i}=-\partial_i\beta,\quad h_{ij}=-a^2\cdot 2\zeta \delta_{ij}, \quad \pi \to \pi _0(t) + \chi,
\end{equation}
where $h_{\mu\nu}$ are metric perturbations about a spatially flat FLRW background
(mostly negative signature), while $\chi$ is the Galileon perturbation.

In (beyond) Horndeski theory, there are two tensor degrees of freedom and one dynamical
degree of freedom in the scalar sector, whose description depends on the gauge choice.
In this paper we discuss the scalar sector only. In what follows, we calculate the derivative part of quadratic action for $\chi$ by utilizing the DPSV trick and compare the result with the corresponding action in the unitary gauge (see e.g. Refs.~\cite{Kolevatov:2017voe,Kobayashi:2011nu})

It has been noted in Ref.\cite{Kolevatov:2017dze} that the DPSV trick may be problematic.
The constraint variables (lapse and shift perturbations $\alpha$, $\partial_i\beta$) are ignored
in the original DPSV
treatment, as they enter the linearized equations without second derivatives. However, the analysis
of the complete quadratic action for metric and Galileon perturbations shows that this may  not be legitimate, generally speaking.
By varying the quadratic action one obtains constraint equations,
which enable one to
express the constraint variables through $\zeta$ and $\chi$:
\begin{equation}
\label{constraints}
\begin{aligned}
\alpha &= a_1\:\dot{\zeta} + a_2\: \dot{\chi} + \dots, \\
\overrightarrow{\nabla}^2 \beta &=  a_3 \overrightarrow{\nabla}^2
\zeta+a_4 \overrightarrow{\nabla}^2 \chi +\dots,
\end{aligned}
\end{equation}
where dots denote the terms with lower derivatives and the coefficients $a_i$ are
combinations
of the Lagrangian functions and their derivatives.
According to the constraint eqs.~\eqref{constraints},
one has to treat
$h_{00} \propto \alpha$ and $h_{0i} \propto \partial_i\beta$ as first-derivative quantities,
and $\dot{\alpha}$ and $\overrightarrow{\nabla}^2\beta$ as second-derivative objects.
So, omitting terms with $\dot{h}_{00}$ and $\partial_i h_{0j}$
in the linearized equations,
as originally done in ~\cite{Deffayet:2010qz}, might be incorrect.

The key observation of Ref.\cite{Kolevatov:2017dze} is that
{\it in Horndeski theory},
there are relations
\begin{equation}
\label{a_i}
a_1 = - a_3, \quad a_2 = - a_4 \; .
\end{equation}
These imply the relation between lapse and shift perturbations, valid
to the leading order in derivatives in the general Horndeski theory (but, as we
see below,
not in beyond Horndeski):
\begin{equation}
\label{ab}
\alpha = - \dot{\beta}.
\end{equation}
The relation~\eqref{ab} enables one to choose the comoving gauge $\beta=0$,
and automatically obtain $\alpha = 0$ (i.e., the gauge turns out to be synchronous as well).
Thus, dropping $\dot{\alpha}$ and $\overrightarrow{\nabla}\beta$ within  the DPSV approach
is, in fact,
legitimate in Horndeski theory and amounts to a specific gauge choice.

This trick has been applied to the $\mathcal{L}_3 + \mathcal{L}_4$ theory
in Ref.\cite{Kolevatov:2017dze}
and has given
the correct result for the derivative part of the quadratic action for $\chi$.
The peculiarity of $\mathcal{L}_3 + \mathcal{L}_4$ theory as compared to the original
$\mathcal{L}_3$ case is that now the trick works for the linearized equations only and
requires a spatially flat FLRW background geometry and a homogeneous background Galileon field,
while originally it worked at a fully non-linear level. In this paper we apply the DPSV approach
to the theory including  $\mathcal{L}_5$  in order to complete the analysis
for the most general
Horndeski theory.

The question that is central for this paper is whether the DPSV trick works in the beyond Horndeski case. As we have already announced above, in beyond Horndeski the trick gives incorrect result for the quadratic action even though it still removes the higher derivatives of metric perturbations.
The former property is due to the fact that in beyond Horndeski theory, relations~\eqref{a_i} are no
longer valid. Thus, $\dot{\alpha}$ and $\overrightarrow{\nabla}\beta$ cannot be set equal to zero
simultaneously by the gauge choice. Thus, the DPSV trick no longer works in the same way as it did in the most general Horndeski theory. However, we suggest the way to modify the trick so that it gives the correct quadratic action for $\chi$ in beyond Horndeski case. 
We show explicitly that by choosing a comoving gauge
and making use of the constraints~\eqref{constraints} one restores the correct result, which agrees with the one in the unitary gauge (see Ref.~\cite{Kolevatov:2017voe}).
 The trick
now involves the calculation of terms
with the {\it first} derivatives of the constraint variables
$h_{00}$ and $h_{0i}$ in the Galileon equation and the use of
the constraint equations to get rid of these variables. Hence, in beyond Horndeski case the DPSV approach looses its advantage in comparison with, for instance, the unitary gauge method, since the modified DPSV approach now forces one to calculate a quadratic action before implementing the trick itself.

\section{DPSV trick for $\mathcal{L}_5$}
\label{sec:DPSV_for_L5}

We consider the DPSV trick in the most general Horndeski theory with the
Lagrangian
$\mathcal{L} = \mathcal{L}_3 + \mathcal{L}_4 + \mathcal{L}_5$ given
by~\eqref{L3}-\eqref{L5}.
The Galileon field equation and Einstein equations have the following form, respectively:
\be
\label{galeq_G5}
& R~\left(\Omega^{(1)}+\Pi^{(1)}\right) + R_{\mu\nu}\left(\Omega^{(2)\mu\nu}+\Pi^{(2)\mu\nu}\right) + R_{\mu\nu\lambda\rho}\left(\Omega^{(3)\mu\nu\lambda\rho} +\Pi^{(3)\mu\nu\lambda\rho}\right) \\
& +2 R_{\mu}^{\rho} R_{\nu\rho} \nabla^{\mu}\pi \nabla^{\nu}\pi G_{5X} -
 R_{\mu\nu} R~ \nabla^{\mu}\pi \nabla^{\nu}\pi G_{5X} +  2 R^{\rho\sigma} R_{\mu\rho \nu \sigma} \nabla^{\mu}\pi \nabla^{\nu}\pi G_{5X} + \dots = 0,
 \ee
\be
\label{einst_G5}
{\cal E}_{\mu \nu} =& R~\left( \Omega^{(4)}_{\mu\nu}+\Pi^{(4)}_{\mu\nu}\right) + R_{\mu\nu}\left(\Omega^{(5)} + \Pi^{(5)}\right) + R_{\mu}^{\rho}\left(\Omega^{(6)}_{\rho\nu} + \Pi^{(6)}_{\rho\nu}\right) + R_{\nu}^{\rho}\left(\Omega^{(7)}_{\rho\mu}+ \Pi^{(7)}_{\rho\mu}\right) \\+ &R_{\rho\sigma}\left(\Omega^{(8)\rho\sigma}_{\mu\nu} + \Pi^{(8)\rho\sigma}_{\mu\nu}\right) +
 R_{\mu\rho\nu\sigma}\left(\Omega^{(9)^{\rho\sigma} }+ \Pi^{(9)\rho\sigma}\right) + R_{\mu\rho\sigma\lambda} \Pi^{(10)\rho\sigma\lambda}_{\nu} \\+
 & R_{\nu\rho\sigma\lambda} \Pi^{(11)\rho\sigma\lambda}_{\mu} + R_{\rho\sigma\lambda\kappa} \Pi^{(12)\rho\sigma\lambda\kappa}_{\mu\nu} + \dots = 0,
\ee
where dots denote the terms, which do not involve the
second derivatives of the {\it metric}, but do involve the second
derivatives of the {\it Galileon}.
The explicit expressions for $\Omega^{(i)}$, $ \Pi^{(i)}$
are given in Appendix A.

In full analogy with the original $\mathcal{L}_3$ case,
the Galileon field equation~\eqref{galeq_G5}
contains second derivatives of both metric and Galileon field. The DPSV approach
makes use of the
Einstein equations~\eqref{einst_G5} in order to get rid of the
second derivatives of metric perturbations, so that there is only one scalar degree of freedom left (with second derivatives)
in the resulting Galileon equation. The
$\mathcal{L}_3+\mathcal{L}_4$ case was proven to admit the trick
at the level of the linearized equations, and for the
background geometry
of a spatially flat FLRW type and a homogeneous background Galileon field.
Let us see that
the same holds for the most general Horndeski theory, including $\mathcal{L}_5$.

Assuming the FLRW type of the background metric and a homogeneous
background Galileon field
$\pi= \pi(t)$, the linearized Galileon equation
reduces to \footnote{In what follows Latin indices take values $1$, $2$, $3$.}
\begin{equation}
\label{galeq_G5_pert}
\begin{aligned}
&g^{ij}  R^{{\bf \{ 1\}}}_{i0j0} \left(M^{(1)}_{L3+L4} +N^{(1)}_{L5}\right) +g^{ij} g^{mn} R^{{\bf \{ 1 \}}}_{imjn} \left(M^{(2)}_{L3+L4} +N^{(2)}_{L5}\right) + Z = 0,
\end{aligned}
\end{equation}
where the
superscript ${\bf \{ 1 \}}$ denotes
linear in perturbations terms of the corresponding objects,
and $Z$ contains
the second derivatives
of $\chi$ but not metric perturbations
(we omit the first and zeroth order derivatives of both metric
and Galileon perturbations).
The explicit forms
of the coefficients $M^{(i)}_{L3+L4}$, $N^{(i)}_{L5}$ and the object
$Z$ are given in Appendix A.
In accordance with the DPSV procedure,
the components of the linearized
Riemann tensor $g^{ij} g^{mn} R^{{\bf \{ 1 \}}}_{imjn}$ and $g^{ij}  R^{{\bf \{ 1 \}}}_{i0j0}$ in~\eqref{galeq_G5_pert}
are to be expressed through the Galileon perturbations. We make use of the
$00-\mbox{component}$ ${\cal E}_{00}^{\{1\}}$
and trace  $\left({\cal E}_{\mu\nu}g^{\mu\nu}\right)^{\{1\}}$ of the linearized Einstein equations:
\begin{subequations}
\label{einst_comp_G5}
\begin{align}
\label{einst_00_G5}
&g^{ij} g^{mn} R^{{\bf \{ 1 \}}}_{imjn} \left(M^{(3)}_{L3+L4} + N^{(3)}_{L5}\right) + Y_{00} = 0,\\
\label{einst_trace_G5}
& g^{ij}  R^{{\bf \{ 1 \}}}_{i0j0} \left(M^{(4)}_{L3+L4} + N^{(4)}_{L5}\right) + g^{ij} g^{mn} R^{{\bf \{ 1 \}}}_{imjn} \left(M^{(5)}_{L3+L4} + N^{(5)}_{L5}\right) + g^{\mu\nu} Y_{\mu\nu} = 0,
\end{align}
\end{subequations}
where $Y_{00} $ and $g^{\mu\nu}Y_{\mu\nu}$ again
involve the second derivatives of the Galileon perturbation, but not
metric perturbations. Their explicit forms are given
in Appendix A.

According to Ref.\cite{Kolevatov:2017dze},
the subtlety with the lapse and shift perturbations
can be safely ignored, since in the general Horndeski case
a comoving gauge is automatically the synchronous one.
Hence, we impose this gauge
and safely drop both $\dot{\alpha}$ and $\overrightarrow{\nabla}\beta$.

The rest of the
procedure is the same as in the $\mathcal{L}_3+\mathcal{L}_4$
case~\cite{Kolevatov:2017dze}.
To obtain the equation that is free of the second derivatives of metric perturbations,
we make use of
eqs.~\eqref{einst_00_G5}  and~\eqref{einst_trace_G5}
and express the structures $g^{ij} g^{mn} R^{{\bf \{ 1 \}}}_{imjn}$
and $g^{ij}  R^{{\bf \{ 1 \}}}_{i0j0}$
in terms of the Galileon perturbation. We
substitute them into  eq.~\eqref{galeq_G5_pert}, which then reads:
\begin{equation}
\label{galeq_after_DPSV}
\begin{aligned}
&\left(\frac{Y_{00}\left(M^{(5)}_{L3+L4} + N^{(5)}_{L5}\right)}{\left(M^{(3)}_{L3+L4} + N^{(3)}_{L5}\right)\left(M^{(4)}_{L3+L4} + N^{(4)}_{L5}\right)} - \frac{g^{\mu\nu}Y_{\mu\nu}}{M^{(4)}_{L3+L4} + N^{(4)}_{L5}}\right) \left(M^{(1)}_{L3+L4} +N^{(1)}_{L5}\right) \\-
& \left(M^{(2)}_{L3+L4} +N^{(2)}_{L5}\right)\frac{Y_{00}}{M^{(3)}_{L3+L4} + N^{(3)}_{L5}} + Z = 0.
\end{aligned}
\end{equation}
The only scalar degree of freedom left
in the linearized equation~\eqref{galeq_after_DPSV} is $\chi$:
at the second derivative level, this equation has the form
$2{\cal D}\chi = 0$, where ${\cal D}$ is a second order differential operator.
The corresponding action is $\int d^4x \,
\chi {\cal D} \chi$. Upon integrating by
parts and using the explicit forms of the objects entering
eq.~\eqref{galeq_after_DPSV},
we reconstruct the derivative part of the
quadratic action for $\chi$:
\begin{equation}
\label{vikman_action_H}
S^{(2)}_{gal} = \int \mathrm{d}t\,\mathrm{d}^3x\,a^3\left(\tilde{\mathcal{A}} \dot{\chi}^2 - \tilde{\mathcal{B}}\dfrac{(\overrightarrow{\nabla}\chi)^2}{a^2}\right),
\end{equation}
where the coefficients $\tilde{\mathcal{A}}$ and $\tilde{\mathcal{B}} $ have the following form
\begin{equation}
\label{AB}
\begin{aligned}
\tilde{\mathcal{A}} =\dfrac{\Sigma\mathcal{G_T}+3 \Theta^2}{\mathcal{G_T} \dot{\pi}^2},\quad \tilde{\mathcal{B}} = \dfrac{\Theta^2}{\mathcal{G_T}^2 \dot{\pi}^2}\left(\dfrac{1}{a}\dfrac{\mathrm{d}}{\mathrm{d}t}\left[\dfrac{a\mathcal{G_T}^2}{\Theta}\right]-\mathcal{F_T}\right).
\end{aligned}
\end{equation}
The explicit expressions for
$\Theta$, $\Sigma$, $\mathcal{G_T}$ and $\mathcal{F_T}$
are given in Appendix A; they coincide with
the expressions given in Ref.~\cite{Kobayashi:2011nu}.
Modulo an overall positive
factor $\Theta^{-2}\mathcal{G_T}^2\dot{\pi}^2$ in the integrand, the action~\eqref{vikman_action_H}
coincides with that found in
Ref.~\cite{Kobayashi:2011nu} in the unitary gauge. This shows that the DPSV approach leads to the correct
quadratic action at the derivative level for the only scalar degree of freedom.
This completes the analysis
of the DPSV approach in the most general Horndeski theory.

\section{Attempting DPSV trick in beyond Horndeski theory}
\label{sec:DPSV_for_BH}

In this Section we show that the situation with the DPSV approach
is tricky in the beyond Horndeski case.
(i) By making use of the Einstein equations,
one {\it can} get rid of the second order and higher
derivatives
of metric perturbations in the Galileon equation (as before,
in the FLRW setting).
(ii) Off hand, this procedure might induce
higher derivative terms in the Galileon equation, but {\it it does not}.
So, the resulting Galileon equation is second order in derivatives of
$\chi$ and at most first order in derivatives of $h_{\mu \nu}$.
These two properties are analogous to those in Horndeski theory.
(iii) However,
disregarding metric perturbations altogether in the resulting
Galileon equation gives {\it wrong} action (at one-derivative level)
for the
scalar degree of freedom. This is in contrast to Horndeski theory.

All these properties are non-trivial in the beyond Horndeski case;
we derive them and give reasons behind the properties
(i) and (ii)
in this Section. Section~\ref{sec:alpha_beta_for_BH} gives the reason behind
the property (iii) and proposes the way to find the correct
one-derivative action in a modified DPSV approach.

To simplify formulas,
we consider the theory with
$  \mathcal{L}_5 = 0$, i.e., with
\begin{equation}
\label{lagrangian_F4}
\mathcal{L} = \mathcal{L}_3+\mathcal{L}_4+\mathcal{L_{BH}}.
\end{equation}
The omitted terms do not introduce any new features. Also,
we keep only the term with $F_4$ in $\mathcal{L_{BH}}$,
eq.~\eqref{BH}. Again, the term with $F_5$ does not give anything new.
Even with this simplification, many formulas are very lengthy,
so we do not write explicitly some of the expressions we meet.
Instead, we try to make clear our steps and discuss the
peculiar features appearing at each step.
We think that omitting lengthy formulas is adequate here, since the outcome
of the entire analysis suggests that the DPSV approach, in fact, is
not the best way to
study the beyond Horndeski theory.

Let us try to follow as close as possible the analysis
done in the Horndeski case, and, in the first place,
get rid of
higher derivatives of metric perturbations in the Galileon equation
by utilizing the Einstein equations.

The Galileon equation for the action \eqref{lagrangian_F4} has
the following form:
\be
\label{Galileon_eq_F4}
& R~\Omega^{(1)} + R_{\mu\nu}\left(\Omega^{(2)\mu\nu} + \Psi^{(2)\mu\nu}\right) + R_{\mu\nu\lambda\rho}\left(\Omega^{(3)\mu\nu\lambda\rho} +\Psi^{(3)\mu\nu\lambda\rho}\right) \\
&-  \nabla^{\mu} \pi \nabla_{\mu} \pi \nabla^{\nu} \pi \left( \nabla_{\nu} R \right)\; F_4 +
2  \nabla^{\mu} \pi \nabla^{\nu} \pi \nabla^{\rho} \pi \left( \nabla_{\rho}R_{\mu\nu} \right)F_4 +\dots = 0,
\ee
where dots again stand for the terms without the second
and
higher derivatives of the metric (but with the second derivatives of Galileon), $\Omega^{(i)}$ are the same as in
eq.~\eqref{galeq_G5},
and the expressions
for $\Psi^{(i)}$ are given in Appendix B. What is new in comparison
with the Horndeski
case is the appearance of $ \nabla_{\rho}R_{\mu\nu} $ and $ \nabla_{\nu} R$, which contain the
third derivatives of the metric. The latter fact is a general feature of the beyond Horndeski theory,
as we alluded to above.

Following the standard procedure, one makes use of the Einstein equations.
Their part involving the second (and higher, if any) derivatives of the metric
reads:
\be
\label{Einstein_eq_F4}
      {\cal E}_{\mu \nu}=&
      R~\Omega^{(4)}_{\mu\nu}+ R_{\mu\nu}\Omega^{(5)} + R_{\mu}^{\rho}\left(\Omega^{(6)}_{\rho\nu} + \Psi^{(6)}_{\rho\nu}\right) + R_{\nu}^{\rho}\left(\Omega^{(7)}_{\rho\mu}+ \Psi^{(7)}_{\rho\mu}\right) \\+ &R^{\rho\sigma}\left(\Omega^{(8)}_{\mu\nu\rho\sigma} + \Psi^{(8)}_{\mu\nu\rho\sigma}\right) +
 R_{\mu\rho\nu\sigma} \Omega^{(9)^{\rho\sigma} } +2  \nabla^{\lambda}\pi \nabla_{\lambda}\pi  \nabla^{\rho}\pi (\nabla_{\rho}\nabla_{\mu}\nabla_{\nu}\pi )F_4 \\-
 &2 \nabla_{\nu}\pi\nabla^{\lambda}\pi\nabla^{\rho}\pi (\nabla_{\rho}\nabla_{\lambda}\nabla_{\mu}\pi) F_4 - 2 \nabla_{\mu}\pi \nabla^{\lambda}\pi \nabla^{\rho}\pi (\nabla_{\rho}\nabla_{\lambda}\nabla_{\nu}\pi) F_4 \\+
& 2 g_{\mu\nu} \nabla^{\lambda}\pi \nabla^{\sigma}\pi \nabla^{\rho}\pi (\nabla_{\rho}\nabla_{\sigma}\nabla_{\lambda}\pi) F_4 + 2 \nabla_{\mu}\pi \nabla_{\nu}\pi \nabla^{\rho}\pi (\nabla_{\sigma}\nabla^{\sigma}\nabla_{\rho}\pi) F_4 \\-
& 2 g_{\mu\nu} \nabla_{\rho}\pi \nabla^{\rho}\pi \nabla^{\sigma}\pi (\nabla_{\lambda}\nabla^{\lambda}\nabla_{\sigma}\pi) F_4
+ \dots = 0.
\ee
One observes that, unlike the Galileon equation~\eqref{Galileon_eq_F4},
the Einstein equations~\eqref{Einstein_eq_F4} do not contain the third derivatives of the metric.
Thus, in beyond Horndeski case the DPSV trick requires extra equations, which involve the
third derivatives of the metric. The way out is to apply a covariant derivative
to the Einstein equations (i.e., consider equations
$\nabla_{\rho} {\cal E}_{\mu\nu} = 0$, which we call the derivative Einstein
equations in what follows) and extract the third derivatives
of the metric from there.
We are going to see that ${\cal E}_{\mu \nu}$ and
$\nabla_{\rho}{\cal E}_{\mu \nu}$ enable one to eliminate the
second and higher derivatives of the metric from the linearized
Galileon equation,
provided the background
geometry is flat FLRW and the background Galileon is homogeneous.

Note that the derivative Einstein equations $\nabla_{\rho} {\cal E}_{\mu\nu} = 0$
involve
the fourth derivatives of the Galileon. So, the procedure just outlined
might introduce the fourth and third derivatives of the Galileon
into the Galileon equation.
We will see, however, that these fourth and third
derivative terms cancel out in the final linearized Galileon equation, so
this linearized equation is second order and describes the single
dynamical degree of freedom in the scalar sector. This is in full
accordance with the general idea behind the beyond Horndeski theory.

As before we drop the first and zeroth derivatives of both metric and Galileon
perturbations in all
linearized equations. This means, in particular, that we erroneously
assume that
there is a gauge which is simultaneously
comoving and synchronous, so that we can drop both $\dot{\alpha}$ and
$\overrightarrow{\nabla}^2\beta$ terms  in a standard manner. We will see
that this is actually {\it not} the case in the beyond Horndeski theory,
but let us proceed with this wrong assumption for the time being.

With a homogeneous and isotropic background, the linearized Galileon
equation is
\begin{equation}
\label{gal_eq_pert_BH}
\begin{aligned}
&g^{ij} g^{mn} R^{{\bf \{ 1 \}}}_{imjn} \left(M^{(2)}_{L3+L4}  - 2 F_{4} H \dot{\pi}^3  \right)
+ g^{ij}  R^{{\bf \{ 1 \}}}_{i0j0} \left(M^{(1)}_{L3+L4}  + 20 F_4 H \dot{\pi}^3 + 8 F_{4X} H \dot{\pi}^5\right) \\-
& \left[\nabla_0 \left(g^{ij} g^{mn} R_{imjn}\right)\right]^{{\bf \{ 1 \}}} F_4 \dot{\pi}^3 + \tilde{Z} = 0,
\end{aligned}
\end{equation}
where $\tilde{Z}$ involves the second derivatives of $\chi$ only.
$00-\mbox{component}$ and trace of the linearized Einstein equations read:
\begin{subequations}
\label{einst_comp_BH}
\begin{align}
\label{einst_00_BH}
{\cal E}_{00}^{\{1\}} = &g^{ij} g^{mn} R^{{\bf \{ 1 \}}}_{imjn}\: M^{(3)}_{L3+L4}  + \tilde{Y}_{00} = 0,\\
\label{einst_trace_BH}
\left(g^{\mu \nu}{\cal E}_{\mu \nu}\right)^{\{1\}} =
& g^{ij}  R^{{\bf \{ 1 \}}}_{i0j0} \: M^{(4)}_{L3+L4}  + g^{ij} g^{mn} R^{{\bf \{ 1 \}}}_{imjn} \: M^{(5)}_{L3+L4} \nonumber\\- &4 F_{4} \dot{\pi}^3 \left( \left(\nabla_{0} \nabla_{\rho} \nabla^{\rho}\pi\right)^{{\bf \{ 1 \}}} -  \left(\nabla_{0} \nabla_{0} \nabla_{0}\pi\right)^{{\bf \{ 1 \}}} \right) + g^{\mu\nu} \tilde{Y}_{\mu\nu} = 0,
\end{align}
\end{subequations}
where $\tilde{Y}_{\mu\nu}$ and its component $\tilde{Y}_{00}$ again
contain the second derivatives of $\chi$ only. Note that the terms like
$\left(\nabla_{0} \nabla_{0} \nabla_{0}\pi\right)^{{\bf \{ 1 \}}}$ contain the
second derivatives of the metric
and the third derivatives of
the Galileon perturbations. Interestingly, the beyond Horndeski term
$\mathcal{L_{BH}}$ contributes only to $\tilde{Y}_{00}$
in eq.~\eqref{einst_00_BH}, but not to the part in
\eqref{einst_00_BH}
with the second derivatives of the metric.

There is only one term in eq.~\eqref{gal_eq_pert_BH}, which involves the third derivatives
of metric perturbations, namely $\left[\nabla_0 \left(g^{ij} g^{mn} R_{imjn}\right)\right]^{{\bf \{ 1 \}}}$.
Hence, in the derivative Einstein equations, we  keep only
$\nabla_0 {\cal E}_{00}$ component (more precisely,
$\nabla_{0}\left(\dot{\pi}^2 {\cal E}_{00}\right)=0$),
which contains this structure:
\begin{equation}
\label{derivative_einst}
\begin{aligned}
&g^{ij} g^{mn} R^{{\bf \{ 1 \}}}_{imjn} \:M^{(6)}_{L3+L4} +
\left[\nabla_0 \left(g^{ij} g^{mn} R_{imjn}\right)\right]^{{\bf \{ 1 \}}}  \dot{\pi}^2 M^{(7)}_{L3+L4}+ \tilde{W} =0,
\end{aligned}
\end{equation}
where $\tilde{W}$ involves not only the second derivatives of both metric and Galileon perturbations,
but also the third derivatives of $\chi$. Let us note that the second derivatives of metric perturbations
in $\tilde{W}$ do not get contracted into Riemann tensors, so we leave them as they are
and check whether
they cancel out, in the final linearized Galileon equation,
with the non-contracted terms
from
$\left(\nabla \nabla \nabla \pi\right)^{{\bf \{ 1 \}}}$
in~\eqref{einst_trace_BH}.
The fourth derivatives of the Galileon, which are present in general,
do not appear in eq.~\eqref{derivative_einst}.

Finally, we express
$\;g^{ij}  R^{{\bf \{ 1 \}}}_{i0j0}$, $\;g^{ij} g^{mn} R^{{\bf \{ 1 \}}}_{imjn}$ and $\;\left[\nabla_0 \left(g^{ij} g^{mn} R_{imjn}\right)\right]^{{\bf \{ 1 \}}}$
from eqs.~\eqref{einst_00_BH}, \eqref{einst_trace_BH}
and \eqref{derivative_einst}
and substitute them into eq.~\eqref{gal_eq_pert_BH}. The resulting
linearized Galileon equation is very cumbersome, and we do not write
it here explicitly. Let us instead describe its main features.
First,
it does not contain the second derivatives of metric perturbations.
This is not so trivial: as we have mentioned, the
second order derivatives of metric perturbations
appear not only through
the components of Riemann tensor, but also through the terms like
$\left(\nabla \nabla \nabla \pi\right)^{{\bf \{ 1 \}}}$.
Second, the resulting
linearized Galileon equation
contains only the second derivatives of the Galileon perturbation:
the third derivatives of the Galileon perturbation indeed cancel out.
Let us note that the cancellations of the second and third derivatives
of metric
perturbations have been verified in an independent way, namely,
we have checked that in the linearized eq.~\eqref{gal_eq_pert_BH}
the terms that cancel out upon the trick are indeed a linear
combination of the corresponding terms in the
linearized eqs.~\eqref{einst_comp_BH} and~\eqref{derivative_einst},
while the third derivatives of the Galileon perturbation
do not get introduced.
So, at this point the analogy with the Hornedski theory appears perfect:
we have obtained the linearized Galileon equation
involving the second derivatives of the Galileon perturbation only.

Having found the second order linearized equation for the Galileon,
$2{\cal D}\chi = 0$, where ${\cal D}$ is a second order differential operator,
we now boldly
reconstruct the quadratic
action  $\int d^4x \,
\chi {\cal D} \chi$. It has the same structure as~\eqref{vikman_action_H}, but with
new functions
$\tilde{\mathcal{A}} (t)$ and $\tilde{\mathcal{B}} (t)$, which we
again do not write explicitly. The key point is to compare them with
the corresponding coefficients
in the unitary gauge, calculated in Ref.\cite{Kolevatov:2017voe}.
As we already announced, we found that
the result of the
DPSV trick differs from that obtained in the unitary gauge:
the coefficients $\tilde{\mathcal{A}}$ and $\tilde{\mathcal{B}}$ are {\it not}
equal to the respective unitary gauge coefficients (modulo an overall
factor).

The reason behind this discrepancy is as follows.
As
noticed in Ref.\cite{Kolevatov:2017dze},
apart from the second derivatives of the Galileon perturbation, the
resulting Galileon equation involves the first derivatives of the metric,
including $\dot{\alpha}$ and $\overrightarrow{\nabla}^2\beta$.
Invariance under the gauge transformations
\begin{equation}
\label{gauge_chi_alpha_beta_zeta}
\chi \to
\chi + \xi_0\dot{\pi},\quad \zeta \to \zeta + \xi_0\dfrac{\dot{a}}{a},\quad \alpha \to \alpha + \dot{\xi_0},\quad \beta \to \beta - \xi_0,
\end{equation}
dictates the following form of this equation (in
both Horndeski and beyond Horndeski theories):
\begin{equation}
\label{GalEq_general_form}
\mathcal{Q}\left(\ddot{\chi} - \dot{\alpha}\dot{\pi}\right) - \mathcal{P}\left(\overrightarrow{\nabla}^2\chi + \overrightarrow{\nabla}^2\beta\dot{\pi}\right) = 0,
\end{equation}
where $\mathcal{Q}$ and $\mathcal{P}$ are expressed in terms of the Lagrangian functions. Now,
even though the terms with $\dot{\alpha}$ and $ \overrightarrow{\nabla}^2\beta$
are formally first order in derivatives,
they
should be treated as second-derivative quantities because of the constraint
eqs.~\eqref{constraints}.
In the general Horndeski theory, both these
terms could be gauged away simultaneously, but, as we see in
the next Section, this is not the case
in the beyond Horndeski theory. Here, the existence of the extra term
with $\dot{\alpha}$ in eq.~\eqref{GalEq_general_form}
in a comoving gauge $\beta=0$ ruins the naive DPSV approach.
On the other hand, keeping this term and expressing it through
$\partial^2 \chi$ restores the agreement with the unitary gauge.

\section{Lapse and shift perturbations in beyond Horndeski}
\label{sec:alpha_beta_for_BH}

To show that $\alpha$ and $\beta$ cannot be gauged away simultaneously in beyond Horndeski
theory, let us consider the quadratic action for the most general beyond Horndeski theory~\eqref{lagrangian}:
\begin{equation}
\label{perturb_action_general_BH}
\begin{aligned}
S^{(2)}_{gr+gal} = \int \mathrm{d}t\,\mathrm{d}^3x\,a^3 \Bigg (
&A_1\:\dot{\zeta}^2 + A_2 \:\dfrac{(\overrightarrow{\nabla}\zeta)^2}{a^2} + A_3\: \alpha^2 + A_4\: \alpha\dfrac{\overrightarrow{\nabla}^2\beta}{a^2} + A_5\: \dot{\zeta}\dfrac{\overrightarrow{\nabla}^2\beta}{a^2} + A_6\: \alpha\dot{\zeta} \\
+A_7&\: \alpha\dfrac{\overrightarrow{\nabla}^2\zeta}{a^2}
+ A_8\:\alpha\dfrac{\overrightarrow{\nabla}^2\chi}{a^2} + A_9\: \dot{\chi}\dfrac{\overrightarrow{\nabla}^2\beta}{a^2} + A_{10}\:\chi\ddot{\zeta} + A_{11}\:\alpha\dot{\chi}  \\
+A_{12}&\:\chi\dfrac{\overrightarrow{\nabla}^2\beta}{a^2}
+ A_{13}\:\chi\dfrac{\overrightarrow{\nabla}^2\zeta}{a^2}
+A_{14}\:\dot{\chi}^2 + A_{15}\:\dfrac{(\overrightarrow{\nabla}\chi)^2}{a^2}  + B_{16}\:\dot{\chi}\dfrac{\overrightarrow{\nabla}^2\zeta}{a^2} \Bigg),
\end{aligned}
\end{equation}
where the coefficients $A_i$ and $B_i$ are expressed through the Lagrangian functions
(see Appendix~B
for the explicit expressions), and the terms which do not contribute to
the higher derivative terms in the corresponding field equations are omitted.
The terms with the coefficients $A_i$ in the quadratic
action
exist in both Horndeski and beyond Horndeski theories.
On the other hand, $B_{16}$
arises only in beyond
Horndeski theories and is crucial in what follows.

Let us find an analog of eqs.~\eqref{a_i} in beyond Horndeski theory.
Following Ref.\cite{Kolevatov:2017dze},
we make use of the invariance of the action~\eqref{perturb_action_general_BH}
and the corresponding linearized equations under the gauge
transformation~\eqref{gauge_chi_alpha_beta_zeta},
and find the constraints on
the coefficients $A_i$ and $B_i$. Let us consider the linearized equations,
following from the action~\eqref{perturb_action_general_BH}, and keep
only the terms
with the highest derivatives of the gauge parameter $\xi_0$ under
the gauge transformation \eqref{gauge_chi_alpha_beta_zeta}.
Varying the action~\eqref{perturb_action_general_BH} with respect to $\beta$ and $\alpha$,
we obtain the following   equations:
\begin{equation}
\label{alpha_and_beta}
\begin{aligned}
A_4\; &\overrightarrow{\nabla}^2\alpha + A_5\:\overrightarrow{\nabla}^2\dot{\zeta} + A_9\:\overrightarrow{\nabla}^2\dot{\chi} + \dots = 0,\\
A_4\; &\overrightarrow{\nabla}^2\beta + A_7\:\overrightarrow{\nabla}^2\zeta + A_8\:\overrightarrow{\nabla}^2\chi + \dots = 0,
\end{aligned}
\end{equation}
where dots denote the terms whose gauge transformation involve lower derivatives of $\xi_0$.
Similarly, the equations obtained by varying the action~\eqref{perturb_action_general_BH}
with respect to $\zeta$ and  $\chi$ are
\begin{equation}
\begin{aligned}
\label{zeta}
&A_7\:\overrightarrow{\nabla}^2\alpha - A_5\: \overrightarrow{\nabla}^2\dot{\beta} + B_{16}\: \overrightarrow{\nabla}^2\dot{\chi}+ \dots = 0,\\
&A_8\:\overrightarrow{\nabla}^2\alpha - A_9\: \overrightarrow{\nabla}^2\dot{\beta} - B_{16}\: \overrightarrow{\nabla}^2\dot{\zeta}+ \dots = 0.
\end{aligned}
\end{equation}
Their gauge invariance implies
\begin{equation}
\label{zeta_and_chi}
A_7 = - A_5 - B_{16} \dot{\pi}, \quad A_8 = - A_9 + B_{16} H.
\end{equation}
Note that there is an extra term $B_{16}$ in eq.~\eqref{zeta_and_chi} as compared to the Horndeski
case, so instead of $\alpha= -\dot{\beta}$, eqs.~\eqref{alpha_and_beta} and \eqref{zeta_and_chi}
lead to the following relation:
\begin{equation}
\label{new_alpha_beta}
\alpha = - \dot{\beta} - \frac{B_{16}}{A_4}\; \left( H\dot{\chi}-\dot{\pi}\dot{\zeta}\right).
\end{equation}
This relation explicitly shows that setting $\beta = 0$ (comoving gauge)
does not
automatically lead to $\alpha = 0$.
Thus, in beyond Horndeski theory, the DPSV formalism
has to be modified by keeping the terms
with $\dot{\alpha}$ and/or $ \overrightarrow{\nabla}^2\beta$ in the linearized
Galileon equation~\eqref{GalEq_general_form}.

This modification is most straightforwardly done in the
comoving gauge $\beta =0$; in this gauge, we should account for non-zero
$\dot{\alpha}$. Let us sketch this modification.
We can get rid of the higher derivatives of metric perturbations in the linearized
Galileon equation~\eqref{gal_eq_pert_BH} with the help of
eqs.~\eqref{einst_comp_BH}
and~\eqref{derivative_einst} in the same way as in
Sec.~\ref{sec:DPSV_for_BH}. But this time the resulting Galileon
equation~\eqref{GalEq_general_form} still contains $\dot{\alpha}$.
We make use of the constraint eqs.~\eqref{alpha_and_beta} with $\beta =0$
to express $\alpha$
in terms of $\dot{\chi}$:
\begin{equation}
\label{alpha_through_chi}
\alpha = \left(\frac{A_5\;A_8}{A_4\;A_7} -\frac{A_9}{A_4} \right)\;\dot{\chi}.
\end{equation}
Substituting the $\dot{\alpha}$ terms with~\eqref{alpha_through_chi}
in eq.~\eqref{GalEq_general_form},
we obtain the linearized Galileon equation with the second derivatives of $\chi$ only.
The reconstructed quadratic action again has the form~\eqref{vikman_action_H}
and
it is
now in complete agreement with the unitary gauge action
found in Ref.\cite{Kolevatov:2017voe}: modulo an overall factor,
the two Lagrangians coincide.
Hence, in beyond Horndeski
theory the (modified)
DPSV approach gives the correct result, provided that
one keeps
the terms with shift and lapse perturbations and makes
use of the constraint equations.
We think the trick becomes fairly
useless in beyond Horndeski case, since it makes heavy use of
rather cumbersome
constraint equations.
It appears easier to work in the unitary gauge from the very beginning.

\section*{Acknowledgements}
The authors are indebted to V.A. Rubakov for fruitful discussions and thoughtful reading of the manuscript.
The work is partly supported by the RFBR grant 18-01-00461-a and by joint grants 17-51-50051-YaF and 16-51-53034-GFEN-a (S.M.). Authors thank the support by the grants of the Foundation for the Advancement of Theoretical Physics and Mathematics "BASIS" (S.M. and V.V.).

\section*{Appendix A}
In this Appendix we give the explicit expressions of the
coefficients entering the formulas of Section~\ref{sec:DPSV_for_L5}.

Groups of coefficients $\Omega^{(i)}$ and
$\Pi^{(i)}$ enter the Einstein and
Galileon equations~\eqref{galeq_G5}, \eqref{einst_G5}.
These coefficients determine the contributions of
$\mathcal{L}_3+\mathcal{L}_4$ and $\mathcal{L}_5$, respectively:
\begin{subequations}
\begin{align*}
& \Omega^{(1)} = 2 \nabla_{\mu} \nabla^{\mu} \pi \: G_{4X} + 4 \nabla^{\mu}\pi \nabla_{\mu}  \nabla_{\nu} \pi \nabla^{\nu}\pi \: G_{4XX} - \: G_{4\pi} + 2  \nabla_{\mu}\pi \nabla^{\mu}\pi \: G_{4\pi X},\\
& \Omega^{(2)\mu\nu} =  \nabla^{\mu} \pi \nabla^{\nu} \pi \: \left( 2 K_{X} + 8 \nabla_{\rho}\nabla^{\rho}\pi \: G_{4XX} -
8 G_{4\pi X} \right)
- 4 \nabla^{\mu} \nabla^{\nu} \pi \: G_{4X} \\
&- 16 \nabla^{\mu}\pi \nabla^{\nu}\nabla_{\rho}\pi \nabla^{\rho}\pi \: G_{4XX},\\
& \Omega^{(3)\mu\nu\lambda\rho} =  -
8 \nabla^{\mu} \pi \nabla^{\lambda} \pi \nabla^{\nu}\nabla^{\rho}\pi \:G_{4XX},\\
&  \Omega^{(4)}_{\mu\nu} = - g_{\mu\nu} \:G_4 +
2 \nabla_{\mu} \pi \nabla_{\nu} \pi \: G_{4X} ,\\
&  \Omega^{(5)} = 2 \:G_{4},\\
&  \Omega^{(6)}_{\rho\nu} = - 4 \nabla_{\nu}\pi \nabla_{\rho}\pi \: G_{4X},\\
&  \Omega^{(7)}_{\rho\mu} = -4 \nabla_{\mu}\pi \nabla_{\rho}\pi \: G_{4X},\\
&  \Omega^{(8)\rho\sigma}_{\mu\nu} = 4 g_{\mu\nu} \nabla^{\rho}\pi \nabla^{\sigma}\pi \: G_{4X},\\
&  \Omega^{(9){\rho\sigma} }= - 4 \nabla^{\rho}\pi \nabla^{\sigma} \pi \: G_{4X}.
\end{align*}
\end{subequations}
\begin{subequations}
\begin{align*}
&\Pi^{(1)}=\nabla_{\rho}\nabla^{\rho}\pi \nabla_{\lambda}\nabla^{\lambda}\pi \:G_{5X} -
  \nabla_{\lambda}\nabla_{\rho}\pi \nabla^{\lambda}\nabla^{\rho}\pi \:G_{5X}  +
   2 \nabla^{\rho}\pi \nabla_{\lambda}\nabla_{\rho}\pi \nabla^{\lambda}\pi
  \nabla_{\sigma}\nabla^{\sigma}\pi \:G_{5XX}  \\\nonumber
& - 2 \nabla^{\rho}\pi \nabla^{\lambda}\pi \nabla_{\sigma}\nabla_{\lambda}\pi
  \nabla^{\sigma}\nabla_{\rho}\pi \:G_{5XX} -  \nabla_{\rho}\nabla^{\rho}\pi \:G_{5\pi}  +
    \nabla_{\rho}\pi \nabla^{\rho}\pi \nabla_{\lambda}\nabla^{\lambda}\pi \:G_{5\pi X}  \\\nonumber
& - 2 \nabla^{\rho}\pi \nabla_{\lambda}\nabla_{\rho}\pi \nabla^{\lambda}\pi \:G_{5\pi X} -
   \frac{1}{2}~ \nabla_{\rho}\pi \nabla^{\rho}\pi \:G_{5\pi\pi },
   \\
\\\nonumber
&\Pi^{(2)\mu\nu} = -4 \nabla^{\nu}\nabla^{\mu}\pi \nabla_{\rho}\nabla^{\rho}\pi \:G_{5X} +
 4 \nabla_{\rho}\nabla^{\nu}\pi \nabla^{\rho}\nabla^{\mu}\pi \:G_{5X} +
   4 \nabla^{\rho}\pi \nabla^{\sigma}\pi \nabla^{\mu}\nabla_{\rho}\pi
  \nabla^{\nu}\nabla_{\sigma}\pi \:G_{5XX} \\\nonumber
& - 4 \nabla^{\rho}\pi \nabla_{\sigma}\nabla_{\rho}\pi \nabla^{\sigma}\pi
     \nabla^{\nu}\nabla^{\mu}\pi \:G_{5XX} -
 8 \nabla^{\mu}\pi \nabla^{\rho}\pi \nabla^{\nu}\nabla_{\rho}\pi
  \nabla_{\sigma}\nabla^{\sigma}\pi \:G_{5XX}  \\\nonumber
& + 2 \nabla^{\mu}\pi \nabla^{\nu}\pi \nabla_{\rho}\nabla^{\rho}\pi
  \nabla_{\sigma}\nabla^{\sigma}\pi \:G_{5XX} +
 8 \nabla^{\mu}\pi \nabla^{\rho}\pi \nabla_{\sigma}\nabla^{\nu}\pi
     \nabla^{\sigma}\nabla_{\rho}\pi \:G_{5XX}  \\\nonumber
& - 2 \nabla^{\mu}\pi \nabla^{\nu}\pi \nabla_{\sigma}\nabla_{\rho}\pi
  \nabla^{\sigma}\nabla^{\rho}\pi \:G_{5XX} -
   2 \nabla_{\rho}\pi \nabla^{\rho}\pi \nabla^{\nu}\nabla^{\mu}\pi \:G_{5\pi X}  +
 8 \nabla^{\mu}\pi \nabla^{\rho}\pi \nabla^{\nu}\nabla_{\rho}\pi \:G_{5\pi X}  \\\nonumber
& - 4 \nabla^{\mu}\pi \nabla^{\nu}\pi \nabla_{\rho}\nabla^{\rho}\pi \:G_{5\pi X} +
  \nabla^{\mu}\pi \nabla^{\nu}\pi \:G_{5\pi\pi } +
   2 \nabla^{\nu}\nabla^{\mu}\pi \:G_{5\pi },
  \\
%
%
%
\end{align*}
\end{subequations}
\begin{subequations}
\begin{align*}
\\\nonumber
& \Pi^{(3)\mu\nu\lambda\rho} = 2 \nabla^{\lambda}\nabla^{\mu}\pi
  \nabla^{\rho} \nabla^{\nu}\pi \:G_{5X} -
 2 \nabla^{\mu}\pi \nabla^{\rho} \nabla^{\lambda}\nabla^{\nu}\pi \:G_{5X} +
 4 \nabla^{\mu}\pi \nabla^{\lambda}\pi \nabla^{\nu}\nabla^{\sigma}\pi
  \nabla^{\rho} \nabla_{\sigma}\pi \:G_{5XX}  \\\nonumber
&  - 4 \nabla^{\mu}\pi \nabla^{\lambda}\pi
  \nabla_{\sigma}\nabla^{\sigma}\pi \nabla^{\rho} \nabla^{\nu}\pi \:G_{5XX} +
   8 \nabla^{\sigma}\pi \nabla^{\mu}\pi
  \nabla^{\lambda}\nabla_{\sigma}\pi \nabla^{\rho} \nabla^{\nu}\pi \:G_{5XX} \\\nonumber
  &+
   4 \nabla^{\mu}\pi \nabla^{\lambda}\pi
  \nabla^{\rho} \nabla^{\nu}\pi \:G_{5\pi X},
  %
  %
\\\nonumber
& \Pi^{(4)}_{\mu\nu}= -\nabla_ {\mu}\pi \nabla_{\nu}\pi \:G_{5\pi } +
   \nabla_ {\mu}\pi \nabla_{\nu}\pi
  \nabla_{\rho} \nabla^{\rho} \pi  \:G_{5X} + \frac{1}{2} g_{ {\mu}{\nu}}
  \nabla_{\rho} \pi \nabla^{\rho} \pi \:G_{5\pi } -
 \nabla_{\nu}\pi \nabla_{\rho} \nabla_ {\mu}\pi \nabla^{\rho} \pi \:G_{5X} \\\nonumber
 & -
 \nabla_ {\mu}\pi \nabla_{\rho} \nabla_{\nu}\pi \nabla^{\rho} \pi \:G_{5X}  +
    g_{ {\mu}{\nu}} \nabla^{\rho} \pi \nabla_{\sigma}\nabla_{\rho} \pi
  \nabla^{\sigma}\pi \:G_{5X},
  \\
  %
  %
  %
\\\nonumber
& \Pi^{(5)}= - \nabla_{\rho} \pi \nabla^{\rho} \pi \:G_{5\pi }  - 2  \nabla^{\rho} \pi \nabla_{\sigma}\nabla_{\rho} \pi \nabla^{\sigma}\pi \:G_{5X},
\\
\\
%
%
%
& \Pi^{(6)}_{\rho\nu}=  2 \nabla_{\nu}\pi \nabla_{\rho} \pi \:G_{5\pi }  + 2 \nabla_{\nu}\pi \nabla^{\sigma}\pi \nabla_{\rho} \nabla_{\sigma}\pi \:G_{5X} - 2 \nabla_{\nu}\pi \nabla_{\rho} \pi \nabla_{\sigma}\nabla^{\sigma}\pi \:G_{5X} \\\nonumber
&+ 2 \nabla_{\rho} \pi \nabla_{\sigma}\nabla_{\nu}\pi \nabla^{\sigma}\pi \:G_{5X},
\\
%
%
%
\\\nonumber
& \Pi^{(7)}_{\rho\mu}= 2 \nabla_{\mu}\pi \nabla_{\rho} \pi \:G_{5\pi } +2  \nabla_{\mu}\pi \nabla^{\sigma}\pi \nabla_{\rho} \nabla_{\sigma}\pi \:G_{5X} - 2  \nabla_{\mu}\pi \nabla_{\rho} \pi \nabla_{\sigma}\nabla^{\sigma}\pi \:G_{5X} \\\nonumber
&+ 2  \nabla_{\rho} \pi \nabla_{\sigma}\nabla_{\mu}\pi \nabla^{\sigma}\pi \:G_{5X},
\\
%
%
\\\nonumber
& \Pi^{(8)\rho\sigma}_{\mu\nu}= 2  \nabla_{\nu}\pi \nabla^{\rho}\pi \nabla^{\sigma}\nabla_{\mu}\pi \:G_{5X} +2  \nabla_{\mu}\pi \nabla^{\rho}\pi \nabla^{\sigma}\nabla_{\nu}\pi \:G_{5X} -2  \nabla_{\mu}\pi \nabla_{\nu}\pi \nabla^{\sigma}\nabla^{\rho}\pi \:G_{5X}
\\\nonumber
& - 2  g_{{\mu}{\nu}} \nabla^{\rho}\pi \nabla^{\sigma}\pi \:G_{5\pi }  - 2  \nabla_{\nu}\nabla_{\mu}\pi \nabla^{\rho}\pi \nabla^{\sigma}\pi \:G_{5X}
 - 4  g_{{\mu}{\nu}} \nabla^{\rho}\pi \nabla^{\lambda}\pi\nabla^{\sigma}\nabla_{\lambda}\pi \:G_{5X} \\\nonumber
 &+ 2  g_{{\mu}{\nu}} \nabla^{\rho}\pi \nabla^{\sigma}\pi\nabla_{\lambda}\nabla^{\lambda}\pi \:G_{5X},
\\
%
%
%
\\\nonumber
& \Pi^{(9)\rho\sigma}=  2 \nabla^{\rho} \pi \nabla^{\sigma}\pi \:G_{5\pi } - 2 \nabla^{\rho} \pi \nabla^{\sigma}\pi\nabla_ {\lambda}\nabla^ {\lambda}\pi \:G_{5X} + 2 \nabla^ {\lambda}\pi \nabla^{\rho} \pi\nabla^{\sigma}\nabla_ {\lambda}\pi \:G_{5X} \\\nonumber
&+ 2 \nabla^ {\lambda}\pi \nabla^{\sigma}\pi\nabla^{\rho} \nabla_ {\lambda}\pi \:G_{5X},
%
\\\nonumber
& \Pi^{(10)\rho\sigma{\lambda}}_{\nu}= - 2 \nabla^{\rho} \pi \nabla^{\sigma}\pi\nabla^{\lambda}\nabla_{\nu}\pi  \:G_{5X}+ 2 \nabla_{\nu}\pi\nabla^{\sigma}\pi\nabla^{\lambda}\nabla^{\rho} \pi  \:G_{5X},
%
\\\nonumber
& \Pi^{(11)\rho\sigma{\lambda}}_{\mu}= - 2 \nabla^{\rho} \pi \nabla^ {\sigma}\pi \nabla_{\lambda}\nabla_{\mu}\pi\:G_{5X} + 2 \nabla_{\mu}\pi \nabla^ {\sigma}\pi\nabla^{\lambda}\nabla^{\rho} \pi \:G_{5X} ,
\\\nonumber
& \Pi^{(12)\rho\sigma\lambda\kappa}_{\mu\nu}= - 2 g_{\mu\nu} \nabla^{\rho}\pi \nabla^{\lambda}\pi\nabla^{\kappa}\nabla^{\sigma}\pi \:G_{5X},
\end{align*}
\end{subequations}

The coefficients $M^{(i)}$ and $N^{(i)}$ appear in the linearized
equations~\eqref{galeq_G5_pert}, \eqref{einst_comp_G5}
and correspond to the
contributions of different subclasses of Horndeski theory,
$\mathcal{L}_3+\mathcal{L}_4$ and $\mathcal{L}_5$, respectively:
\begin{subequations}
\begin{align*}
&M^{(1)}_{L3+L4} =-2\left[G_{4\pi} - 4G_{4X} H \dot{\pi} + (2 G_{4\pi X} - K_{X})\dot{\pi}^2 - 8 G_{4XX} H \dot{\pi}^3 \right]  ,\\
&M^{(2)}_{L3+L4} =-G_{4\pi} + 2 G_{4X} H \dot{\pi} + 2 G_{4X} \ddot{\pi} + 2 \dot{\pi}^2 (G_{4\pi X} + 2 G_{4XX} \ddot{\pi}),\\
\end{align*}
\end{subequations}
\begin{subequations}
\begin{align*}
&M^{(3)}_{L3+L4} = (2 G_{4X} \dot{\pi}^2 - G_{4} ),\\
&M^{(4)}_{L3+L4} = 4 M^{(3)}_{L3+L4},\\
&M^{(5)}_{L3+L4} = 2 (G_{4X} \dot{\pi}^2 - G_{4}),\\
&M^{(6)}_{L3+L4} = -\dot{\pi} \left[ 2 G_4 \ddot{\pi} + \dot{\pi}^2 \left(G_{4\pi} - 6 G_{4X} \ddot{{\pi}}\right) - 2 \dot{\pi}^4 \left(G_{4\pi X} + 2 G_{4XX}  \ddot{{\pi}}\right)\right],
\end{align*}
\end{subequations}
\begin{subequations}
\begin{align*}
&M^{(7)}_{L3+L4} = M^{(3)}_{L3+L4},\\
&N^{(1)}_{L5} =-4 G_{5\pi} H \dot{\pi} + 6 G_{5X} H^2 \dot{\pi}^2 + 4 G_{5XX} H^2 \dot{\pi}^4 - 4 G_{5\pi X} H \dot{\pi}^3,\\
&N^{(2)}_{L5} =G_{5\pi X} H \dot{\pi}^3 - G_{5\pi} H \dot{\pi} -G_{5\pi} \ddot{\pi} + 2 G_{5X} H \dot{\pi} \ddot{\pi} + G_{5X} \dot{\pi}^2\frac{\ddot{a}}{a}
-G_{5\pi X} \dot{\pi}^2 \ddot{\pi} +\\
&+2 G_{5XX} H \dot{\pi}^3 \ddot{\pi}- \frac12 G_{5\pi\pi} \dot{\pi}^2,\\
&N^{(3)}_{L5} = - \frac12 \dot{\pi}^2 \left(G_{5\pi} - 2 G_{5X} H \dot{\pi}\right) ,\\
&N^{(4)}_{L5} = 4 N^{(3)}_{L5},\\
&N^{(5)}_{L5} = G_{5X}  \dot{\pi}^2 \left(H \dot{\pi} + \ddot{\pi}\right).
\end{align*}
\end{subequations}

The objects entering the linearized equations
\eqref{galeq_G5_pert} and \eqref{einst_comp_G5} are
\begin{subequations}
  \begin{align*}
    & Z=2\ddot{\chi} \big[-F_{X} - 2 F_{XX} \dot{\pi}^2 + K_{\pi } - 6 K_{X} H
\dot{\pi} + K_{\pi X} \dot{\pi}^2 - 6 K_{XX}H \dot{\pi}^3 - 6 G_{4X}
H^2 \\
&+ 18 G_{4\pi X} H \dot{\pi} - 48 G_{4XX} H^2 \dot{\pi}^2 + 12
G_{4\pi XX} H \dot{\pi}^3 - 24 G_{4XXX} H^2 \dot{\pi}^4 + 3 G_{5\pi }
H^2 \\
&- 6 G_{5X} H^3 \dot{\pi} + 15 G_{5\pi X} H^2 \dot{\pi}^2  - 14
G_{5XX} H^3 \dot{\pi}^3 + 6 G_{5\pi XX} H^2 \dot{\pi}^4 - 4 G_{5XXX}
H^3 \dot{\pi}^5\big] \\
 & + 2\frac{\overrightarrow{\nabla}^2\chi}{a^2} \big[F_{X} - K_{\pi } + 2
K_{X} \ddot{\pi} + 4 K_{X} H \dot{\pi} + K_{\pi X} \dot{\pi}^2 + 2
K_{XX} \dot{\pi}^2 \ddot{\pi} + 2 G_{4X} H^2 \\
&+ 4 G_{4X}
\frac{\ddot{a}}{a} - 12 G_{4\pi X} H \dot{\pi} - 6 G_{4\pi X}
\ddot{\pi} + 12 G_{4XX} H^2 \dot{\pi}^2 + 24 G_{4XX} H \dot{\pi}
\ddot{\pi} + 8 G_{4XX} \dot{\pi}^2 \frac{\ddot{a}}{a} \\
&-2 G_{4\pi \pi X}
\dot{\pi}^2 + 8 G_{4\pi XX} H \dot{\pi}^3 - 4 G_{4\pi XX} \dot{\pi}^2
\ddot{\pi} + 16 G_{4XXX} H \dot{\pi}^3 \ddot{\pi} - G_{5\pi } H^2 - 2
G_{5\pi } \frac{\ddot{a}}{a} \\
&+ 2 G_{5X} H^2 \ddot{\pi} + 4 G_{5X} H
\dot{\pi} \frac{\ddot{a}}{a} - 8 G_{5\pi X} H \dot{\pi} \ddot{\pi} - 3
G_{5\pi X} H^2 \dot{\pi}^2 - 2 G_{5\pi X} \dot{\pi}^2
\frac{\ddot{a}}{a} + 10 G_{5XX} H^2 \dot{\pi}^2 \ddot{\pi} \\
&+ 4 G_{5XX}
H \dot{\pi}^3 \frac{\ddot{a}}{a} - 2 G_{5\pi \pi X} H \dot{\pi}^3 - 4
G_{5\pi XX} H \dot{\pi}^3 \ddot{\pi} + 2 G_{5\pi XX} H^2 \dot{\pi}^4 +
4 G_{5XXX} H^2 \dot{\pi}^4 \ddot{\pi}\big],
    \\
    \\
\end{align*}
\end{subequations}
\begin{subequations}
\begin{align*}
    & Y_{00} = 2 \frac{\overrightarrow{\nabla}^2\chi}{a^2} \big[K_{X} \dot{\pi}^4 -
G_{4\pi} \dot{\pi}^2 + 4 G_{4X} H \dot{\pi}^3  - 2 G_{4\pi X} \dot{\pi}^4 +
8 G_{4XX} H \dot{\pi}^5 - 2 G_{5\pi} H \dot{\pi}^3 \\
&+ 3 G_{5X} H^2
\dot{\pi}^4 - 2 G_{5\pi X} H \dot{\pi}^5 + 2 G_{5XX} H^2 \dot{\pi}^6\big],
\\
\\
    & g^{\mu \nu}Y_{\mu \nu} =2 \ddot{\chi} \big[- 3 K_{X} \dot{\pi}^2 + 3 G_{4\pi} -12 G_{4X} H
\dot{\pi} + 6 G_{4\pi X} \dot{\pi}^2  - 24 G_{4XX} H \dot{\pi}^3 + 6
G_{5\pi} H \dot{\pi} \\
&- 9 G_{5X} H^2 \dot{\pi}^2  + 6 G_{5\pi X} H
\dot{\pi}^3 - 6 G_{5XX} H^2 \dot{\pi}^4\big] + 2\frac{\overrightarrow{\nabla}^2 \chi}{a^2}
   \big[K_{X} \dot{\pi}^2 - 3 G_{4\pi} + 8 G_{4X} H \dot{\pi} \\
&+ 4 G_{4X}
\ddot{\pi} + 2 G_{4\pi X} \dot{\pi}^2 + 8 G_{4XX} H \dot{\pi}^3 + 8
G_{4XX}\ddot{\pi} \dot{\pi}^2 - 2 G_{5\pi} \ddot{\pi}- 4 G_{5\pi} H
\dot{\pi} + 4 G_{5X} H \ddot{\pi} \dot{\pi}\\
& + 3 G_{5X} H^2 \dot{\pi}^2
+ 2 G_{5X} \frac{\ddot{a}}{a} \dot{\pi}^2 - G_{5\pi\pi} \dot{\pi}^2 - 2
G_{5\pi X}\ddot{\pi} \dot{\pi}^2   + 2 G_{5XX} H^2 \dot{\pi}^4 + 4
G_{5XX} H \dot{\pi}^3 \ddot{\pi}\big].
  \end{align*}
  \end{subequations}

The quantities that are involved in the coefficients $\mathcal{A}$ and $\mathcal{B}$ in ~\eqref{AB} have the following form:
\begin{subequations}
    \begin{align*}
    &\mathcal{G_T}=2G_4 - 4G_{4X}\dot{\pi}^2 + G_{5\pi}\dot{\pi}^2 - 2HG_{5X}\dot{\pi}^3,\\
    &\mathcal{F_T} =2G_4 - 2G_{5X}\dot{\pi}^2\ddot{\pi}-G_{5\pi}\dot{\pi}^2,\\
    &\Theta=-K_X\dot{\pi}^3+2G_4H-8HG_{4X}\dot{\pi}^2-8HG_{4XX}\dot{\pi}^4+G_{4\pi}\dot{\pi}+2G_{4\pi X}\dot{\pi}^3-
    \\\nonumber&-5H^2G_{5X}\dot{\pi}^3-2H^2G_{5XX}\dot{\pi}^5+3HG_{5\pi}\dot{\pi}^2+2HG_{5\pi X}\dot{\pi}^4,\\
    \label{eq:Sigma_coeff_setup}
    &\Sigma=F_X\dot{\pi}^2+2F_{XX}\dot{\pi}^4+12HK_X\dot{\pi}^3+6HK_{XX}\dot{\pi}^5-K_{\pi}\dot{\pi}^2-K_{\pi X}\dot{\pi}^4-
    \\\nonumber&-6H^2G_4+42H^2G_{4X}\dot{\pi}^2+96H^2G_{4XX}\dot{\pi}^4+24H^2G_{4XXX}\dot{\pi}^6-
    \\\nonumber&-6HG_{4\pi}\dot{\pi}-30HG_{4\pi X}\dot{\pi}^3-12HG_{4\pi XX}\dot{\pi}^5+30H^3G_{5X}\dot{\pi}^3+
    \\\nonumber&+26H^3G_{5XX}\dot{\pi}^5+4H^3G_{5XXX}\dot{\pi}^7-18H^2G_{5\pi}\dot{\pi}^2-27H^2G_{5\pi X}\dot{\pi}^4-
    \\\nonumber&-6H^2G_{5\pi XX}\dot{\pi}^6.
    \end{align*}
\end{subequations}
These are the same as in Ref.~\cite{Kobayashi:2011nu}.

\section*{Appendix B}

Here we collect the explicit expressions occurring in beyond Horndeski
theory as discussed in Secs.~\ref{sec:DPSV_for_BH} and
\ref{sec:alpha_beta_for_BH}.

We begin with the coefficients entering eqs.~\eqref{Galileon_eq_F4},
\eqref{Einstein_eq_F4}. We write them in the case $F_5=0$ and $F_4 \neq 0$
considered in  Sec.~\ref{sec:DPSV_for_BH}:
\begin{subequations}
\begin{align*}
& \Psi^{(2)\mu\nu} =  \nabla^{\mu} \pi \nabla^{\nu} \pi \: \left[ 10 \nabla_{\rho}\nabla^{\rho}\pi F_4 + 4 \nabla^{\sigma}\pi \nabla_{\sigma}\pi \nabla_{\rho}\nabla^{\rho}\pi F_{4X} + 4 \nabla^{\rho}\pi \nabla_{\rho}  \nabla_{\sigma} \pi \nabla^{\sigma}\pi F_{4X} \right] -\\-
& 2 \nabla^{\mu} \nabla^{\nu} \pi \:\nabla_{\rho}\pi \nabla^{\rho}\pi F_4 -  \nabla^{\mu}\pi \nabla^{\nu}\nabla_{\rho}\pi \nabla^{\rho}\pi \: \left[8 F_4 +8 \nabla^{\lambda} \pi \nabla_{\lambda} \pi F_{4X}\right],\\
& \Psi^{(3)\mu\nu\lambda\rho} = - \nabla^{\mu} \pi \nabla^{\lambda} \pi \nabla^{\nu}\nabla^{\rho}\pi \: \left( 8 F_4 + 4 \nabla^{\sigma} \pi \nabla_{\sigma} \pi F_{4X} \right), \\
&  \Psi^{(6)}_{\rho\nu} = - 2 \nabla_{\nu}\pi \nabla_{\rho}\pi \nabla^{\lambda}\pi \nabla_{\lambda}\pi\: F_4,\\
&  \Psi^{(7)}_{\rho\mu} = -2 \nabla_{\mu}\pi \nabla_{\rho}\pi \nabla^{\lambda}\pi \nabla_{\lambda}\pi \:F_4,\\
& \Psi^{(8)}_{\mu\nu\rho\sigma} = \nabla_{\rho}\pi \nabla_{\sigma}\pi \: \left( 2 g_{\mu\nu} \nabla^{\lambda}\pi \nabla_{\lambda}\pi F_4 + 2 \nabla_{\mu}\pi \nabla_{\nu}\pi F_4\right).
\end{align*}
\end{subequations}

Finally, we give the expressions for the
coefficients $A_i$ and $B_i$ entering
the quadratic
action~\eqref{perturb_action_general_BH} in the most general beyond Horndeski theory~\eqref{lagrangian}:
\\
\begin{subequations}
\begin{align*}
&A_1=3\left[-2G_4+4G_{4X}\dot{\pi}^2-G_{5\pi}\dot{\pi}^2+2HG_{5X}\dot{\pi}^3 + 2 F_{4}\dot{\pi}^4\ + 6 H F_{5}\dot{\pi}^5\right],\\
&A_2=2G_4-2G_{5X}\dot{\pi}^2\ddot{\pi}-G_{5\pi}\dot{\pi}^2,\\
&A_3=F_X\dot{\pi}^2+2F_{XX}\dot{\pi}^4+12HK_X\dot{\pi}^3+6HK_{XX}\dot{\pi}^5-K_{\pi}\dot{\pi}^2-K_{\pi X}\dot{\pi}^4\\
\nonumber&-6H^2G_4+42H^2G_{4X}\dot{\pi}^2+96H^2G_{4XX}\dot{\pi}^4+24H^2G_{4XXX}\dot{\pi}^6\\
\nonumber&-6HG_{4\pi}\dot{\pi}-30HG_{4\pi X}\dot{\pi}^3-12HG_{4\pi XX}\dot{\pi}^5+30H^3G_{5X}\dot{\pi}^3\\
\nonumber&+26H^3G_{5XX}\dot{\pi}^5+4H^3G_{5XXX}\dot{\pi}^7-18H^2G_{5\pi}\dot{\pi}^2-27H^2G_{5\pi X}\dot{\pi}^4\\
\nonumber&-6H^2G_{5\pi XX}\dot{\pi}^6+90H^2F_4\dot{\pi}^4+78H^2F_{4X}\dot{\pi}^6+12H^2F_{4XX}\dot{\pi}^8\\
\nonumber&+168H^3F_5\dot{\pi}^5+102H^3F_{5X}\dot{\pi}^7+12H^3F_{5XX}\dot{\pi}^9,\\
&A_4=2\big[K_X\dot{\pi}^3-2G_4H+8HG_{4X}\dot{\pi}^2+8HG_{4XX}\dot{\pi}^4-G_{4\pi}\dot{\pi}-2G_{4\pi X}\dot{\pi}^3\\
\nonumber&+5H^2G_{5X}\dot{\pi}^3+2H^2G_{5XX}\dot{\pi}^5-3HG_{5\pi}\dot{\pi}^2-2HG_{5\pi X}\dot{\pi}^4\\
\nonumber&+10HF_4\dot{\pi}^4+4HF_{4X}\dot{\pi}^6+21H^2F_5\dot{\pi}^5+6H^2F_{5X}\dot{\pi}^7\big],\\
&A_5=-\dfrac{2}{3}A_1,\\
&A_6=-3A_4,\\
&A_7 = -A_5 -B_{16}\dot{\pi},\\
&A_8=2\big[K_X\dot{\pi}^2-G_{4\pi}-2G_{4\pi X}\dot{\pi}^2+4 HG_{4X}\dot{\pi}+8HG_{4XX}\dot{\pi}^3-2HG_{5\pi}\dot{\pi}\\
&-2HG_{5\pi X}\dot{\pi}^3+3 H^2G_{5X}\dot{\pi}^2+2 H^2G_{5XX}\dot{\pi}^4 + 10 H F_{4}\dot{\pi}^3
+ 4 H F_{4X}\dot{\pi}^5 \\
&+ 21 H^2 F_{5}\dot{\pi}^4 + 6 H^2 F_{5X}\dot{\pi}^6\big],\\
&A_9=- \big(A_8 - B_{16} H\big),\\
&A_{10}=-3\big(A_8-B_{16} H\big) ,\\
\end{align*}
\end{subequations}
\begin{subequations}
\begin{align*}
&A_{11}=2\big[-F_X\dot{\pi}-2F_{XX}\dot{\pi}^3+K_\pi\dot{\pi}-6HK_{XX}\dot{\pi}^4-9HK_X\dot{\pi}^2+K_{\pi X}\dot{\pi}^3\\
&+3HG_{4\pi}+24HG_{4\pi X}\dot{\pi}^2+12H G_{4\pi XX}\dot{\pi}^4-18H^2G_{4X}\dot{\pi}-72H^2G_{4XX}\dot{\pi}^3\\
&-24H^2G_{4XXX}\dot{\pi}^5+9H^2G_{5\pi}\dot{\pi}+21H^2G_{5\pi X}\dot{\pi}^3+6H^2G_{5\pi XX}\dot{\pi}^5\\
&-15H^3G_{5X}\dot{\pi}^2-
20H^3G_{5XX}\dot{\pi}^4-4H^3G_{5XXX}\dot{\pi}^6 - 60 H^2 F_{4}\dot{\pi}^3 - 66 H^2 F_{4X}\dot{\pi}^5\\
& - 12 H^2 F_{4XX}\dot{\pi}^7 - 105 H^3 F_{5}\dot{\pi}^4 - 84 H^3 F_{5X}\dot{\pi}^6 - 12 H^3 F_{5XX} \dot{\pi}^8
\big],\\
&A_{12}=2\big[F_X\dot{\pi}-K_\pi\dot{\pi}+3HK_X\dot{\pi}^2-HG_{4\pi}+G_{4\pi\pi}\dot{\pi}-10HG_{4\pi X}\dot{\pi}^2+6H^2G_{4X}\dot{\pi}\\
&+12H^2G_{4XX}\dot{\pi}^3-3H^2G_{5\pi}\dot{\pi}+HG_{5\pi\pi}\dot{\pi}^2-4H^2G_{5\pi X}\dot{\pi}^3+3H^3G_{5X}\dot{\pi}^2\\
&+2H^3G_{5XX}\dot{\pi}^4 + 12 H^2 F_{4}\dot{\pi}^3 + 6 H^2 F_{4X}\dot{\pi}^5 - 2 H F_{4\pi}\dot{\pi}^4 + 15 H^3 F_{5} \dot{\pi}^4 + 6 H^3 F_{5X} \dot{\pi}^6 \\
&- 3 H^2 F_{5\pi} \dot{\pi}^5
\big],\\
&A_{13}= 2\big[4HG_{4X}\dot{\pi} + 4G_{4X}\ddot{\pi} + 8 G_{4XX}\dot{\pi}^2\ddot{\pi} - 2G_{4\pi} +4 G_{4\pi X}\dot{\pi}^2 + 2 H^2G_{5X}\dot{\pi}^2\\
&+2 \dot{H}G_{5X}\dot{\pi}^2+4 H G_{5X}\dot{\pi}\ddot{\pi}+4HG_{5XX}\dot{\pi}^3\ddot{\pi} -2HG_{5\pi}\dot{\pi}-2G_{5\pi}\ddot{\pi}+2HG_{5\pi X}\dot{\pi}^3\\
&-2G_{5\pi X}\dot{\pi}^2\ddot{\pi}-G_{5\pi\pi}\dot{\pi}^2 + 2 H F_{4}\dot{\pi}^3 + 6 F_{4} \ddot{\pi}\dot{\pi}^2 +4 F_{4X}\ddot{\pi}\dot{\pi}^4 + 2 F_{4\pi}\dot{\pi}^4 + 24 H F_{5} \ddot{\pi}\dot{\pi}^3\\
&+6 H^2 F_{5} \dot{\pi}^4 + 6 \dot{H} F_{5}\dot{\pi}^4 + 12 H F_{5X}\ddot{\pi}\dot{\pi}^5 + 6 H F_{5\pi} \dot{\pi}^5
\big],\\
&A_{14}=F_{X} + 2 F_{XX}\dot{\pi}^2- K_{\pi} + 6 H K_{X}\dot{\pi} -
 K_{\pi X}\dot{\pi}^2 + 6 H K_{XX}\dot{\pi}^3 + 6 H^2 G_{4 X} \\
 &-18 H G_{4\pi X}\dot{\pi} + 48 H^2 G_{4 XX}\dot{\pi}^2 -
 12 H G_{4\pi XX}\dot{\pi}^3 + 24 H^2 G_{4 XXX}\dot{\pi}^4 +
 6 H^3 G_{5 X}\dot{\pi} \\
 &- 3 H^2 G_{5\pi}-
15 H^2 G_{5\pi X}\dot{\pi}^2 + 14 H^3 G_{5 XX}\dot{\pi}^3 +
 4 H^3 G_{5 XXX}\dot{\pi}^5 - 6 H^2 G_{5\pi XX}\dot{\pi}^4 \\
 &+
 36 H^2 F_{4}\dot{\pi}^2 + 54 H^2 F_{4 X}\dot{\pi}^4 +
 12 H^2 F_{4 XX}\dot{\pi}^6 + 60 H^3 F_{5}\dot{\pi}^3 +
 66 H^3 F_{5 X}\dot{\pi}^5 \\
 &+ 12 H^3 F_{5 XX}\dot{\pi}^7 ,\\
&A_{15}=-F_{X} - 4 H K_{X}\dot{\pi} - 2 K_{X}\ddot{\pi} + K_{\pi} -
 K_{\pi X}\dot{\pi}^2 - 2  K_{XX}\dot{\pi}^2 \ddot{\pi} -
 6 H^2 G_{4 X} \\
 &- 4 \dot{H} G_{4 X} -
 20 H^2 G_{4 XX}\dot{\pi}^2 - 8 \dot{H} G_{4 XX}\dot{\pi}^2 -
 24 H  G_{4 XX}\dot{\pi}\ddot{\pi} + 12 H  G_{4\pi X}\dot{\pi} \\
 &+
 6 G_{4\pi X}\ddot{\pi} - 16 H  G_{4 XXX}\dot{\pi}^3\ddot{\pi} -
 8 H G_{4\pi XX}\dot{\pi}^3 + 4  G_{4\pi XX}\dot{\pi}^2\ddot{\pi} +
 2  G_{4\pi \pi X}\dot{\pi}^2 \\
 &- 4 H^3 G_{5 X}\dot{\pi} -
 4 H\dot{H} G_{5 X}\dot{\pi} - 2 H^2  G_{5 X}\ddot{\pi} +
 3 H^2 G_{5\pi} + 2 \dot{H} G_{5\pi} +
 5 H^2  G_{5\pi X}\dot{\pi}^2 \\
 &+
 2 \dot{H} G_{5\pi X}\dot{\pi}^2 +
 8 H  G_{5\pi X}\dot{\pi}\ddot{\pi} -
 4 H^3  G_{5 XX}\dot{\pi}^3 - 4 H \dot{H} G_{5 XX}\dot{\pi}^3 -
 10 H^2  G_{5 XX}\dot{\pi}^2\ddot{\pi} \\
 &-
 4 H^2  G_{5 XXX}\dot{\pi}^4\ddot{\pi} -
 2 H^2  G_{5\pi XX}\dot{\pi}^4 +
 4 H  G_{5\pi XX}\dot{\pi}^3\ddot{\pi} +
 2 H  G_{5\pi\pi X}\dot{\pi}^3 - 20  F_{4} H^2 \dot{\pi}^2 \\
 &-
 10 \dot{H} F_{4}  \dot{\pi}^2 -
 24 H F_{4}\dot{\pi} \ddot{\pi} - 10 H^2  F_{4 X}\dot{\pi}^4 -
 4 \dot{H} F_{4 X}\dot{\pi}^4 -
 36 H  F_{4 X}\dot{\pi}^3\ddot{\pi} - 6 H  F_{4\pi}\dot{\pi}^3 \\
 &-
 8 H  F_{4 XX}\dot{\pi}^5\ddot{\pi} -
 4 H  F_{4\pi X}\dot{\pi}^5 - 30 H^3 F_{5} \dot{\pi}^3 -
 36 H\dot{H} F_{5} \dot{\pi}^3 -
 60 H^2 F_{5} \dot{\pi}^2\ddot{\pi} -
 12 H^3 F_{5 X}\dot{\pi}^5 \\
 &- 12 H \dot{H} F_{5 X}\dot{\pi}^5 -
 66 H^2  F_{5 X}\dot{\pi}^4\ddot{\pi} -
 12 H^2 F_{5\pi}\dot{\pi}^4 -
 12 H^2  F_{5 XX}\dot{\pi}^6\ddot{\pi} -
 6 H^2  F_{5\pi X}\dot{\pi}^6,\\
&B_{16}=4 F_{4} \dot{\pi}^3 + 12 H F_{5}\dot{\pi}^4.
\end{align*}
\end{subequations}
These expressions show explicitly, in particular,
that the relations~\eqref{zeta_and_chi},
derived from the gauge invariance of the field equations, are indeed valid.

\end{document}